\documentclass[conference]{IEEEtran}

\usepackage[T1]{fontenc}
\usepackage[latin9]{inputenc}
\usepackage{float}
\usepackage{amsmath}
\usepackage{amsthm}
\usepackage{amssymb}
\usepackage{graphicx}
\usepackage{stfloats}
\usepackage{cite} 
\usepackage{xcolor}
\usepackage{bbm}
\usepackage{multirow}
\usepackage{multicol}
\usepackage{subfig}

\usepackage{algorithm}
\usepackage{algpseudocode}

\usepackage[letterpaper, 
    top=0.75in, 
    bottom=1.05in, 
    left=0.625in, 
    right=0.625in, 
    columnsep=0.24in
]{geometry}

\makeatletter

\floatstyle{ruled}
\newfloat{algorithm}{tbp}{loa}
\providecommand{\algorithmname}{Algorithm}
\floatname{algorithm}{\protect\algorithmname}

\theoremstyle{plain}

\theoremstyle{plain}

\IEEEoverridecommandlockouts
\usepackage{cite}
\usepackage{amsfonts} 
\usepackage{textcomp}
\usepackage{etoolbox}

\def\BibTeX{{\rm B\kern-.05em{\sc i\kern-.025em b}\kern-.08em
    T\kern-.1667em\lower.7ex\hbox{E}\kern-.125emX}}

\patchcmd{\@maketitle}
  {\addvspace{0.5\baselineskip}\egroup}
  {\addvspace{-1\baselineskip}\egroup}
  {}
  {}

\makeatother

\providecommand{\propositionname}{Proposition}
\providecommand{\theoremname}{Theorem}

\begin{document}

\title{\huge Efficient Alternating Optimization for Hybrid Digital-Wave Beamforming in SIM-Assisted Cell-Free Massive MIMO}

\author{\IEEEauthorblockN{Eunhyuk Park$^{1}$, Seok-Hwan Park$^{1}$, Osvaldo Simeone$^2$, and Marco Di Renzo$^{3,4}$} \IEEEauthorblockA{
$^1$School of Electrical Engineering, Hanyang University, Ansan, Korea\\
$^2$Institute for Intelligent Networked Systems, Northeastern University London, London, U.K. \\
$^3$Center for Telecommun. Research, Department of Engineering, King's College London, London, U.K.\\
$^4$CNRS and CentraleSup\'elec, Institute of Electronics and Digital Technologies (IETR), Rennes, France \\
Email: \{eunhyukpark, seokhwanpark\}@hanyang.ac.kr, o.simeone@northeastern.edu, } marco.di\_renzo@kcl.ac.uk and marco.direnzo@centralesupelec.fr}
\maketitle
\begin{abstract}
Stacked intelligent metasurfaces (SIMs) have recently emerged as a promising architecture for large-scale beamforming systems, including cell-free massive MIMO (CF-mMIMO), due to their cost-effective wave-domain signal processing capabilities.
However, existing algorithms for the joint optimization of digital and SIM-enabled wave-domain beamforming typically incur prohibitive computational complexity. 
In this work, we propose an efficient alternating optimization (AO) algorithm for weighted sum-rate maximization in SIM-assisted CF-mMIMO systems employing hybrid digital-wave beamforming.
Unlike prior approaches that rely on general-purpose optimization solvers or per-element gradient ascent methods, the proposed algorithm updates the digital and wave-domain beamforming variables on a per-access point (AP) or per-SIM-layer basis, enabling closed-form updates at each step.
Numerical results demonstrate that the proposed algorithm reduces the computational complexity by more than 99\% compared to existing algorithms while achieving nearly identical sum-rate performance.
\end{abstract}


\theoremstyle{theorem}
\newtheorem{theorem}{Theorem} 
\theoremstyle{proposition}
\newtheorem{proposition}{Proposition} 
\theoremstyle{lemma}
\newtheorem{lemma}{Lemma} 
\theoremstyle{corollary}
\newtheorem{corollary}{Corollary} 
\theoremstyle{definition}
\newtheorem{definition}{Definition}
\theoremstyle{remark}
\newtheorem{remark}{Remark}

\section{Introduction} \label{sec:intro}

\let\thefootnote\relax\footnotetext{
The work of E. Park and S.-H. Park was partially supported by the National Research Foundation (NRF) of Korea, funded by the MSIT under Grant RS-2026-25468472 and by the Regional Innovation System \& Education (RISE) program through the Jeonbuk RISE Center, funded by the MOE and the Jeonbuk State, Republic of Korea under Grant 2025-RISE-13-JBU.
The work of O. Simeone was supported by the European Research Council (ERC) under the European Union's Horizon Europe Programme (grant agreement No. 101198347), by an Open Fellowship of the Engineering and Physical Sciences Research Council (EPSRC) (EP/W024101/1), and by the EPSRC project EP/X011852/1.
The work of M. Di Renzo was supported in part by the ERC under the European Union's Horizon Europe Programme WePhICom (grant agreement number 101225119), as well as by the European Union through the Horizon Europe project COVER under grant agreement number 101086228, the Horizon Europe project UNITE under grant agreement number 101129618, the Horizon Europe project INSTINCT under grant agreement number 101139161, and the Horizon Europe project TWIN6G under grant agreement number 101182794, as well as by the Agence Nationale de la Recherche (ANR) through the France 2030 project ANR-PEPR Networks of the Future under grants agreementNF-YACARI 22-PEFT-0005, and by the CHIST-ERA project PASSIONATE under grant agreements CHIST-ERA-22-WAI-04 and ANR-23-CHR4-0003-01. Also, the work of M. Di Renzo was supported in part by the EPSRC, part of UK Research and Innovation, and the UK Department of Science, Innovation and Technology through the CHEDDAR Telecom Hub under grant EP/Y037421/1, through the HASC Telecom Hub under grant EP/Y037197/1, and through the TITAN Telecom Hub under grant EP/Y037243/1.}

In cell-free massive multiple-input multiple-output (CF-mMIMO) systems, regarded as a pivotal technology for sixth-generation (6G) wireless networks, a large number of distributed access points (APs) are coordinated by a central processor (CP) to simultaneously serve mobile user equipments (UEs) \cite{Ngo:TWC17}.
By adopting this distributed architecture, the network can eliminate traditional cell boundaries and provide uniform seamless connectivity while mitigating inter-cell interference.
However, the practical implementation of CF-mMIMO faces significant challenges due to high hardware costs and system complexity. Although equipping APs with large-scale antenna arrays can alleviate the need for dense AP deployment by effectively extending per-AP coverage \cite{Xu:JSTSP25}, the requirement of a dedicated radio frequency (RF) chain for each antenna element leads to a prohibitive hardware cost.

To address these challenges, stacked intelligent metasurfaces (SIMs) have recently emerged as a promising low-cost wave-domain processing technology \cite{An:ICC23, An:JSAC23}. An SIM is composed of multiple programmable layers, each containing nearly passive meta-atoms capable of manipulating the phase shifts of incoming electromagnetic waves. 
By enabling beamforming directly in the wave domain, SIMs can significantly reduce the number of RF chains required at each AP.
Consequently, SIMs facilitate the scalable and cost-effective deployment of CF-mMIMO networks.
Recent studies \cite{Hu:TVT25, Shi:TWC25-DL, Park:TWC26, Bahingayi:WCL26} have developed alternating optimization (AO) frameworks to jointly optimize digital and wave-domain beamforming variables.
However, in \cite{Hu:TVT25} and \cite{Shi:TWC25-DL}, the digital processing was limited to power control with fixed beamforming directions. Moreover, the AO algorithm in \cite{Park:TWC26} suffers from high computational complexity due to its reliance on generic optimization solvers or element-wise updates, which limits its practical implementability.
Although \cite{Bahingayi:WCL26} developed efficient algorithms without restricting digital beamforming to power control, it focuses on a single-AP network without inter-AP cooperation.


To bridge the gap between performance and implementation complexity, this work proposes an efficient AO-based framework for the joint optimization of digital and wave-domain beamforming to maximize the weighted sum-rate in SIM-assisted CF-mMIMO systems.
We first apply the matrix Lagrangian duality transformation \cite{Shen:TN19} to reformulate the considered non-convex problem.
To optimize the digital beamforming keeping fixed the wave-domain beamforming, we utilize the reduced-complexity weighted minimum mean squared error (R-WMMSE) method \cite{Zhao:TSP23}, which decomposes the problem into per-AP subproblems, each admitting a closed-form solution.
To optimize the wave-domain beamforming, we partition the wave-domain variables on a per-layer basis rather than a per-AP basis and sequentially optimize the variables of each layer using closed-form solutions.
Specifically, a penalty-based regularization term is incorporated into the objective function to handle the unit-modulus constraint \cite{Park:TWC26}.
Unlike \cite{Park:TWC26}, the proposed framework admits closed-form updates through the AO procedure, leading to a substantially lower computational complexity.
Numerical results demonstrate that the proposed AO algorithm for hybrid digital and wave-domain beamforming achieves nearly identical performance as the optimization approach in \cite{Park:TWC26}, while significantly reducing computational complexity and achieving substantial performance gains compared with alternative moderate-complexity baseline schemes.

\section{System Model\label{sec:System-Model}}

\begin{figure}
\centering\includegraphics[width=0.8\linewidth]{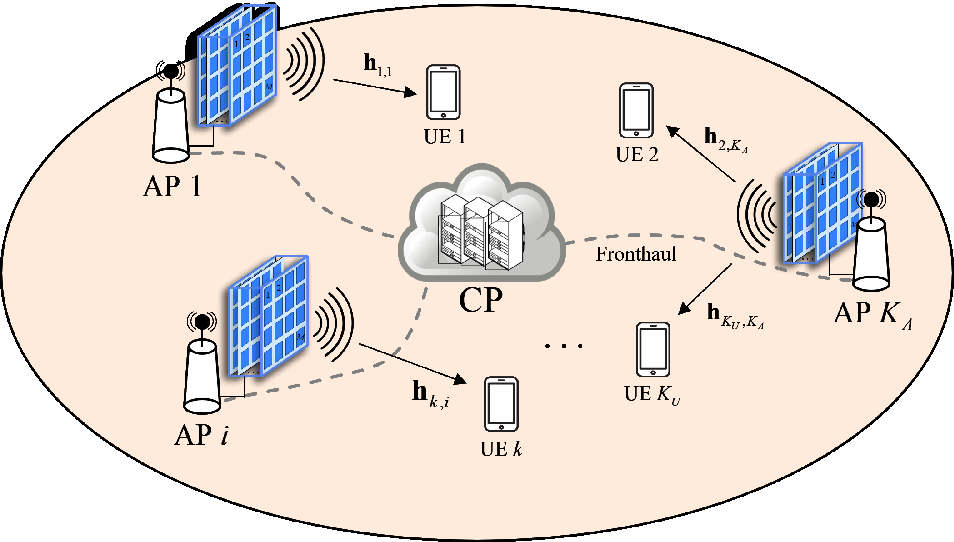}
\caption{\small Considered SIM-assisted CF-mMIMO system.} \label{fig:system-model}
\vspace{-3mm}
\end{figure}

As shown in Fig. \ref{fig:system-model}, we consider the downlink of a CF-mMIMO system comprising a CP, $K_A$ APs, and $K_U$ single-antenna UEs. Each AP is equipped with $N$ antennas, each of them connected to a dedicated RF chain.
Although high-precision massive MIMO beamforming can be achieved with a large number of antennas $N$, deploying many antennas is generally undesirable due to the hardware cost and power consumption of RF chains.
To preserve the beamforming gains of massive MIMO while maintaining a moderate number of antennas $N$, we assume that each AP is equipped with an SIM placed in front of the antenna array.
Each SIM consists of $L$ metasurface layers, with each layer comprising $M$ meta-atoms with $M \gg N$. 
The wave-domain processing at each metasurface layer is described in the following subsections.
For notational convenience, we define the sets $\mathcal{K}_A = \{1,2,\ldots,K_A\}$, $\mathcal{K}_U = \{1,2,\ldots,K_U\}$, $\mathcal{N} = \{1,2,\ldots,N\}$, $\mathcal{L} = \{1,2,\ldots,L\}$, and $\mathcal{M} = \{1,2,\ldots,M\}$.

\subsection{Channel Model \label{sub:CSI-model}}

The channel vector $\mathbf{h}_{k,i}\in\mathbb{C}^{M\times 1}$ between the output metasurface layer of AP $i$ and UE $k$ is modeled as $\mathbf{h}_{k,i} \sim \mathcal{CN} ( \mathbf{0}, \beta_{k,i} \mathbf{R}_{k,i} )$, where $\beta_{k,i}$ denotes the large-scale pathloss between UE $k$ and AP $i$, and $\mathbf{R}_{k,i}$ represents the spatial correlation matrix.
Under the assumption of isotropic scattering and uniformly distributed multipath components, the $(m,m^\prime)$th element of $\mathbf{R}_{k,i}$ is given by $\mathbf{R}_{k,i}(m,m^\prime) = \text{sinc} (2d_{m,m^\prime}^\text{meta}/\lambda)$ \cite{Bjornson:WCL21}, where $\text{sinc}(x) = \text{sin}(\pi x)/(\pi x)$, $d_{m,m^\prime}^\text{meta}$ denotes the spacing between the $m$th and $m^{\prime}$th meta-atoms, and $\lambda$ is the wavelength.

We assume that the CP has an estimate of each channel vector $\mathbf{h}_{k,i}$, which is denoted by $\hat{\mathbf{h}}_{k,i}$, and can be formulated, in terms of $\mathbf{h}_{k,i}$, as
\begin{align}
    \mathbf{h}_{k,i} = \hat{\mathbf{h}}_{k,i} + \mathbf{e}_{k,i}, \label{eq:CSI-model}
\end{align}
where the estimation error vector $\mathbf{e}_{k,i} \sim \mathcal{CN}(\mathbf{0}, \boldsymbol{\Psi}_{k,i})$ is uncorrelated with the estimated channel vector $\hat{\mathbf{h}}_{k,i}$ \cite{Choi:TWC20}.



\subsection{Hybrid Digital- and Wave-Domain Beamforming \label{sub:Downlink-transmission}}

In downlink data transmission, the data signals $\{s_k\}_{k\in\mathcal{K}_U}$ intended for the UEs undergo digital beamforming and wave-domain beamforming, as detailed in this subsection.

\subsubsection{Digital Beamforming} \label{subsub:digital-beamforming-downlink}

The CP performs digital beamforming to the data signals $\{s_k\}_{k\in\mathcal{K}_U}$, thereby generating the precoded signal $\mathbf{x} = [\mathbf{x}_1^H \cdots \mathbf{x}_{K_A}^H]^H \in\mathbb{C}^{N K_A \times 1}$ given by
\begin{align}
    \mathbf{x} = \sum\nolimits_{k\in\mathcal{K}_U} \mathbf{v}_k s_k. \label{eq:digital-beamforming}
\end{align}
Here, $\mathbf{v}_k = [(\mathbf{v}_{k,1})^H \cdots (\mathbf{v}_{k,K_A})^H]^H \in \mathbb{C}^{N K_A \times 1}$ denotes the digital beamforming vector for $s_k$, and the subvectors $\mathbf{x}_i\in\mathbb{C}^{N\times 1}$ and $\mathbf{v}_{k,i}\in\mathbb{C}^{N\times 1}$ are the beamformed signal and beamforming vectors, respectively, associated with AP $i$.

The transmitted signal vector $\mathbf{x}_i$ from AP $i$ is subject to the following constraint:
\begin{align}
    \mathbb{E}\left[\|\mathbf{x}_i\|^2\right] = \sum\nolimits_{k\in\mathcal{K}_U} \|\mathbf{v}_{k,i}\|^2 \leq P_\text{tx}. \label{eq:power-constraint}
\end{align}

\subsubsection{Wave-Domain Beamforming} \label{subsub:wave-beamforming-downlink}

The digital beamformed signal $\mathbf{x}_i$ of AP $i$ is processed by the SIM placed in front of AP $i$, leading to the output signal $\bar{\mathbf{x}}_i\in\mathbb{C}^{M \times 1}$ given by
\begin{align}
    \bar{\mathbf{x}}_i &= \boldsymbol{\Phi}_{i,L} \mathbf{W}_{i,L} \boldsymbol{\Phi}_{i,L-1} \cdots \boldsymbol{\Phi}_{i,2} \mathbf{W}_{i,2} \boldsymbol{\Phi}_{i,1} \mathbf{T}_i \, \mathbf{x}_i, \label{eq:wave-beamforming}
\end{align}
where $\mathbf{T}_i\in\mathbb{C}^{M\times N}$ denotes the transmission matrix from the $N$ antennas to the input metasurface layer, $\mathbf{W}_{i,l}\in\mathbb{C}^{M\times M}$ represents the transmission matrix between the $(l-1)$th and $l$th metasurface layers, and $\boldsymbol{\Phi}_{i,l} = \text{diag}(\{ e^{j\theta_{i,l,m}} \}_{m\in\mathcal{M}}) \in\mathbb{C}^{M\times M}$ is the phase shift matrix of the $l$th metasurface layer. Based on Rayleigh-Sommerfeld diffraction theory \cite{Lin:Sci18}, the $(m,m^\prime)$th element of $\mathbf{W}_{i,l}$ is given by 
\begin{align}
    \mathbf{W}_{i,l}(m, m^\prime) = \frac{S_i d_{i,\text{Layer}}}{d_{i,l,m,m^\prime}^2} \left( \frac{1}{2 \pi d_{i,l,m,m^\prime}} - \frac{j}{\lambda} \right) e^{\frac{j 2 \pi d_{i,l,m,m^\prime}}{\lambda}}, \nonumber
\end{align}
where $S_i$ is the area of each meta-atom, $d_{i,\text{Layer}}$ represents the spacing between adjacent metasurface layers, $d_{i,l,m,m^\prime}$ denotes the transmission distance between the $m^\prime$th meta-atom in the $(l-1)$th layer and the $m^\prime$th meta-atom in the $l$th layer, and $\lambda$ is the wavelength. Similarly, the $(m,m^\prime)$th element of $\mathbf{T}_i$ can be obtained based on the relative positions of the meta-atoms. We define the overall wave-domain processing matrix at AP $i$ as 
\begin{align}
    \mathbf{G}_i = \boldsymbol{\Phi}_{i,L} \mathbf{W}_{i,L} \boldsymbol{\Phi}_{i,L-1} \cdots \boldsymbol{\Phi}_{i,2} \mathbf{W}_{i,2} \boldsymbol{\Phi}_{i,1}. \label{eq:wave-processing-matrix}
\end{align}

\subsubsection{Achievable Rates} \label{subsub:achievable-rate-downlink}

Given the digital beamforming $\mathbf{v}=\{\mathbf{v}_k\}_{k\in\mathcal{K}_U}$ and wave-domain beamforming variables $\boldsymbol{\theta} \!\!=\!\! \{\theta_{i,l,m}\}_{i\in\mathcal{K}_A, l\in\mathcal{L}, m\in\mathcal{M}}$,
the achievable data rate of UE $k$ is
\begin{align}
    R_k = f_k \big( \mathbf{v}, \boldsymbol{\theta} \big) = \log_2\left( 1 + \big| \dot{\mathbf{h}}_k^H \mathbf{v}_k \big|^2 / \text{IF}_k\big(\mathbf{v}, \boldsymbol{\theta}\big)  \right), \label{eq:data-rate-dl}
\end{align}
where $\dot{\mathbf{h}}_k = [\dot{\mathbf{h}}_{k,1}^H \, \cdots \, \dot{\mathbf{h}}_{k,K_A}^H]^H$ with $\dot{\mathbf{h}}_{k,i} = \mathbf{T}_i^H \mathbf{G}_i^H \hat{\mathbf{h}}_{k,i}$, and 
the interference-plus-noise power at UE $k$ is
\begin{align}
    \text{IF}_k \big(\mathbf{v}, \boldsymbol{\theta} \big) = &\sum\nolimits_{k^{\prime}\in\mathcal{K}_U\setminus\{k\}} \big| \dot{\mathbf{h}}_k^H\mathbf{v}_{k^{\prime}} \big|^2 \nonumber \\ &+ \sum\nolimits_{k^{\prime}\in\mathcal{K}_U}   \mathbf{v}_{k^{\prime}}^H\bar{\mathbf{T}}^H\bar{\mathbf{G}}^H \bar{\mathbf{\Psi}}_k \bar{\mathbf{G}}\bar{\mathbf{T}}\mathbf{v}_{k^{\prime}}  + \sigma_z^2, \label{eq:IF}
\end{align} 
using the definition $\bar{\mathbf{G}} = \text{blkdiag}(\{\mathbf{G}_i\}_{i\in\mathcal{K}_A})$, $\bar{\mathbf{T}} = \text{blkdiag}(\{\mathbf{T}_i\}_{i\in\mathcal{K}_{A}})$, $\bar{\mathbf{\Psi}}_k = \text{blkdiag}(\{ \mathbf{\Psi}_{k,i} \}_{i\in\mathcal{K}_A})$. Also, the noise power is $\sigma_z^2$.

\subsection{Problem Definition} \label{sub:problem-downlink}

We aim at maximizing the weighted sum-rate $\sum_{k\in\mathcal{K}_U} \alpha_k R_k$ by jointly optimizing the digital beamforming $\mathbf{v}$ and the wave-domain beamforming $\boldsymbol{\theta}$.
The problem can be formulated as
\begin{subequations} \label{eq:problem-original}
\begin{align}
    \underset{\mathbf{v}, \boldsymbol{\theta}} {\mathrm{max.}}\,\,\, & \sum\nolimits_{k\in\mathcal{K}_U} \alpha_k f_k\big(\mathbf{v}, \boldsymbol{\theta}\big) \, \label{eq:problem-original-objective} \\
 \mathrm{s.t. }\,\,\,\,\, & \sum\nolimits_{k\in\mathcal{K}_U} \! \|\mathbf{v}_{k,i}\|^2 \leq P_\text{tx}, \, \forall i\in\mathcal{K}_A, \label{eq:problem-original-power} \\
 & \theta_{i,l,m} \in [0,2\pi), \, \forall (i,l,m)\in \mathcal{K}_A\times\mathcal{L}\times\mathcal{M}. \label{eq:problem-original-wave-unit-modulus} 
\end{align}
\end{subequations}
The formulated problem was recently tackled in \cite{Park:TWC26} by using an AO algorithm. However, the proposed approach has a high computational complexity.
In this paper, we propose a new and efficient algorithm that significantly reduces the complexity while having a negligible performance loss.

\section{Proposed Efficient Hybrid Beamforming} \label{sec:low-complexity}

In this section, we propose an efficient AO algorithm for hybrid digital and wave-domain beamforming.

\subsection{Optimization of Digital Beamforming} \label{sub:problem-digital-low}

The problem in (\ref{eq:problem-original}) with the wave-domain beamforming $\boldsymbol{\theta}$ fixed remains non-convex due to the objective function in (\ref{eq:problem-original-objective}).
To convert the log-fractional expression in $f_k(\mathbf{v}, \boldsymbol{\theta})$ into a more tractable form, we introduce a lower bound for $f_k(\mathbf{v}, \boldsymbol{\theta})$ using the matrix Lagrangian duality transform \cite[Thm. 2]{Shen:TN19}, as follows:
\begingroup
\allowdisplaybreaks
\begin{align}
    &f_k\big( \mathbf{v}, \boldsymbol{\theta} \big) \geq \tilde{f}_k\big(\mathbf{v}, \boldsymbol{\theta}, \tau_k, \omega_k\big) \nonumber \\
    & = \log_2 \left( 1+\,\tau_k \right) - \frac{\tau_k}{\ln2} + \frac{1+\tau_k}{\ln2} \bigg[ 2\text{Re} \Big\{ \mathbf{v}_k^H \dot{\mathbf{h}}_k \omega_k \Big\} \nonumber \\ &\quad- | \omega_k |^2 \left( \big| \dot{\mathbf{h}}_k^H \mathbf{v}_{k} \big|^2 + \text{IF}_k\big(\mathbf{v}, \boldsymbol{\theta}\big) \right) \bigg], \label{eq:convexified-objective-dl}
\end{align}
\endgroup
for any $\tau_k \in \mathbb{R}_+$ and $\omega_k \in \mathbb{C}$.
The lower bound in (\ref{eq:convexified-objective-dl}) is particularly useful since it becomes a quadratic function of $\mathbf{v}$ when $\boldsymbol{\theta}$ and the auxiliary variables $\{\tau_k,\omega_k\}$ are fixed. Moreover, it is tight when $\tau_k$ and $\omega_k$ are set to
\begingroup
\allowdisplaybreaks
\begin{align}
    &\tau_k = \frac{ \big| 
    \dot{\mathbf{h}}_k^H \mathbf{v}_k \big|^2 } { \text{IF}_k\big(\mathbf{v}, \boldsymbol{\theta}\big)}, \,\,\omega_k = \frac{ \dot{\mathbf{h}}_k^H \mathbf{v}_k }{ \big|\dot{\mathbf{h}}_k^H  \mathbf{v}_k \big|^2 + \text{IF}_k\big(\mathbf{v}, \boldsymbol{\theta}\big) }. \label{eq:opt-auxiliary-tau-omega}
\end{align}
\endgroup
Using the lower bound in (\ref{eq:convexified-objective-dl}), the optimization problem for digital beamforming $\textbf{v}$ can be reformulated as
\begin{align}
    \underset{\mathbf{v}, \boldsymbol{\tau}, \boldsymbol{\omega}} {\mathrm{max.}}\,\,\, \sum\nolimits_{k\in\mathcal{K}_U} \alpha_k \tilde{f}_k\big(\mathbf{v}, \boldsymbol{\theta}, \tau_k, \omega_k\big) \,\,\,\,
 \mathrm{s.t. }\,\, \textrm{(\ref{eq:problem-original-power})}. \label{eq:problem-convexified-digital}
\end{align}
where $\boldsymbol{\tau}=\{\tau_k\}_{k\in\mathcal{K}_U}$ and $\boldsymbol{\omega}=\{\omega_k\}_{k\in\mathcal{K}_U}$. Keeping the auxiliary variables $\{\boldsymbol{\tau},\boldsymbol{\omega}\}$ fixed, the problem over $\mathbf{v}$ is convex. However, obtaining a closed-form solution is difficult and would require the use of general-purpose convex solvers.

To avoid this reliance, we utilize the R-WMMSE approach in \cite{Zhao:TSP23, Yoo:WCL24}, which partitions $\mathbf{v}$ into block variables $\{\mathbf{v}_{A,i}\}_{i\in\mathcal{K}_A}$. Here  we define $\mathbf{v}_{A,i} = [ \mathbf{v}_{1,i}^H \, \dots\, \mathbf{v}_{K_U,i}^H ]^H \in \mathbb{C}^{NK_U\times 1}$, which collects the beamforming vectors associated with AP $i$.
The block variables $\{\mathbf{v}_{A,i}\}_{i\in\mathcal{K}_A}$ are then sequentially optimized in closed form. 
For fixed $\{\mathbf{v}_{A,j}\}_{j\in\mathcal{K}_A\setminus\{i\}}$, the subproblem with respect to $\mathbf{v}_{A,i}$ reduces to the following quadratic convex problem:
\begin{align}
    \!\!\!\!\underset{\mathbf{v}_{A,i}} {\mathrm{min.}}\,\,\, & \mathbf{v}_{A,i}^H \mathbf{Q}_i^\text{D}\mathbf{v}_{A,i} + 2\text{Re}\Big\{ \!\left(\mathbf{b}_i^\text{D}\right)^H\mathbf{v}_{A,i} \Big\} \,\,\,\,
 \mathrm{s.t. }\,\, \textrm{(\ref{eq:problem-original-power})},  \label{eq:problem-digital-low}
\end{align}
where $\mathbf{Q}_i^\text{D} \in \mathbb{C}^{NK_U \times NK_U}$ and $\mathbf{b}_i^\text{D} \in \mathbb{C}^{NK_U \times 1} $ are
\begin{subequations} \label{eq:digital-Q-b}
\begin{align}
    & \mathbf{Q}_i^\text{D} = \frac{1}{\ln2} \sum\nolimits_{k\in\mathcal{K}_U} \alpha_k \left( 1+\tau_k \right) |\omega_k|^2 \left( \mathbf{I}_{K_U} \otimes \mathbf{B}_{k,i}^\text{D} \right), \nonumber \\
    & \mathbf{b}_i^\text{D} = \frac{1}{\ln2} \sum\nolimits_{k\in\mathcal{K}_U} \alpha_k \left( 1+\tau_k \right) \Bigl( |\omega_k|^2 \mathbf{C}_{k,i}^\text{D}  - \omega_k \mathbf{E}_{k}^H \Bigl) \dot{\mathbf{h}}_{k,i}, \nonumber
\end{align}    
\end{subequations}
with $\mathbf{B}_{k,i}^\text{D} = \dot{\mathbf{h}}_{k,i} \dot{\mathbf{h}}_{k,i}^H + \mathbf{T}_i^H \mathbf{G}_i^H \boldsymbol{\Psi}_{k,i}\mathbf{G}_i \mathbf{T}_i$, $\mathbf{C}_{k,i}^\text{D} = \sum\nolimits_{k^\prime \in \mathcal{K}_U}A_{k,k^\prime,i}^* \mathbf{E}_{k^\prime}^H$, $\mathbf{E}_k = \big[ \mathbf{0}_{N\times(k-1)} \, \mathbf{I}_N \, \mathbf{0}_{N\times(K_U-k)} \big]$, and $A_{k,k^\prime,i} = \sum\nolimits_{j \in \mathcal{K}_A \setminus \{i\}} \mathbf{v}_{k^\prime,j}^H \dot{\mathbf{h}}_{k,j}$.

The Lagrangian of the subproblem (\ref{eq:problem-digital-low}) is given by
\begin{align}
    \mathcal{L} \left( \mathbf{v}_{A,i}, \lambda_i \right) = &\mathbf{v}_{A,i}^H \mathbf{Q}_i^\text{D}\mathbf{v}_{A,i} + 2\text{Re}\Big\{ \!\left(\mathbf{b}_i^\text{D}\right)^H\mathbf{v}_{A,i} \Big\}  \nonumber \\ &+ \lambda_i \left( \|\mathbf{v}_{A,i}\|^2 - P_\text{tx} \right), \label{eq:lagrange-digital}
\end{align}
with a Lagrange multiplier $\lambda_i \geq 0$.
By applying the first-order Karush-Kuhn-Tucker (KKT) conditions, the optimal solution is obtained as
\begin{align}
    \mathbf{v}_{A,i} = - \left( \mathbf{Q}_i^\text{D} + \lambda_i^\star \mathbf{I}_{NK_U} \right)^{-1} \mathbf{b}_i^\text{D}, \label{eq:lagrange-digital-opt}
\end{align}
where the optimal multiplier $\lambda_i^\star$ can be efficiently obtained via the bisection method \cite[Sec. 4.2.5]{Boyd:Cambridge04}, exploiting the monotonic decrease of $\|\left( \mathbf{Q}_i^\text{D} + \lambda_i^\star \mathbf{I}_{NK_U} \right)^{-1} \mathbf{b}_i^\text{D}\|^2$ with respect to $\lambda_i$.

As discussed in Sec. \ref{sub:algorithm-complexity}, given $\boldsymbol{\theta}$, the optimization of $\mathbf{v}$ is performed through alternating updates between $\mathbf{v}$ and $\{\boldsymbol{\tau}, \boldsymbol{\omega}\}$.
The convergence of this alternating process is guaranteed by the framework established in \cite{Shen:TN19}.
In particular, $\mathbf{v}$ is updated by sequentially optimizing the per-AP block variables $\mathbf{v}_{A,1}, \ldots, \mathbf{v}_{A,K_A}$.
These updates ensure a monotonic increase of the weighted sum-rate objective, thereby guaranteeing convergence to a stationary point of (\ref{eq:problem-convexified-digital}).

\subsection{Optimization of Wave-Domain Beamforming} \label{sub:problem-wave-low}

For a purpose similar to that in the digital beamforming optimization, we apply the matrix Lagrangian duality transform to optimize the wave-domain beamforming variables $\boldsymbol{\theta}$, as follows:
\begin{align}
    \!\!\!\!\underset{\boldsymbol{\theta}, \boldsymbol{\tau}, \boldsymbol{\omega}}{\mathrm{max.}} \,\,\, & \sum\nolimits_{k\in\mathcal{K}_U} \alpha_k \tilde{f}_k\big(\mathbf{v}, \boldsymbol{\theta}, \tau_k, \omega_k\big) \,\,\,\,
    \textrm{s.t. } \,\, \textrm{(\ref{eq:problem-original-wave-unit-modulus})}. \label{eq:problem-wave-low} 
\end{align}    
However, problem (\ref{eq:problem-wave-low}) is more challenging than (\ref{eq:problem-convexified-digital}) due to the multiplicative coupling among the inter-layer phase shift variables. This strong coupling makes it difficult to jointly optimize all wave-domain variables.
To address this issue, we adopt a sequential optimization strategy over per-layer phase shift variables $\boldsymbol{\varphi}_1, \boldsymbol{\varphi}_2, \ldots, \boldsymbol{\varphi}_L$. Here we have defined a vector $\boldsymbol{\varphi}_l = [ \boldsymbol{\varphi}_{1,l}^H \, \dots\, \boldsymbol{\varphi}_{K_A,l}^H ]^H \in \mathbb{C}^{MK_A \times 1}$, which collects the phase shifts of $l$th layer across all APs, with $\boldsymbol{\varphi}_{i,l} \in \mathbb{C}^{M \times 1}$ containing the diagonal elements of $\boldsymbol{\Phi}_{i,l}$.
For fixed $\{\boldsymbol{\varphi}_{l^{\prime}}\}_{l^{\prime}\in\mathcal{L}\setminus\{l\}}$, the subproblem with respect to $\boldsymbol{\varphi}_{l}$ can be simplified as
\begin{subequations} \label{eq:problem-wave-low-CQ}
\begin{align}
    \!\!\!\!\underset{\boldsymbol{\varphi}_{l}} {\mathrm{min.}}\,\,\, & \boldsymbol{\varphi}_{l}^H \mathbf{Q}_l^\text{W} \boldsymbol{\varphi}_{l} + 2\text{Re}\Big\{ \!\left(\mathbf{b}_l^\text{W}\right)^H \boldsymbol{\varphi}_{l} \Big\} \, \label{eq:problem-wave-objective-low-CQ} \\
 \mathrm{s.t. }\quad & |\boldsymbol{\varphi}_{i,l}(m)| = 1, \forall (i, m) \in \mathcal{K}_A \times \mathcal{M}, \label{eq:problem-wave-unit-modulus-low-CQ} 
\end{align}
\end{subequations}
where $\mathbf{Q}_l^\text{W} \in \mathbb{C}^{MK_A \times MK_A}$ and $\mathbf{b}_l^\text{W} \in \mathbb{C}^{MK_A \times 1} $ are 
\begin{subequations} \label{eq:wave-Q-b}
\begin{align} 
    & \mathbf{Q}_l^\text{W} = \frac{1}{\ln2} \sum\nolimits_{k\in\mathcal{K}_U} \!\!\alpha_k \left( 1+\tau_k \right) |\omega_k|^2 \left( \mathbf{B}_{k,l}^\text{W} \left(\mathbf{B}_{k,l}^{\text{W}}\right)^H \!+\! \mathbf{C}_{k,l}^\text{W} \right)\!, \nonumber \\
    & \mathbf{b}_l^\text{W} = - \frac{1}{\ln2} \sum\nolimits_{k\in\mathcal{K}_U} \!\!\alpha_k \left( 1+\tau_k \right) \omega_k \left\{\! \bigl( \bar{\mathbf{X}}_l^H \hat{\mathbf{h}}_{k} \bigl) \odot \bigl( \bar{\mathbf{Y}}_l \bar{\mathbf{T}} \mathbf{v}_k \bigl)^* \!\right\}\!, \nonumber
\end{align}    
\end{subequations}
with $\mathbf{B}_{k,l}^\text{W} = \text{diag} \left( \bar{\mathbf{X}}_l^H \hat{\mathbf{h}}_k \right) \left( \bar{\mathbf{Y}}_l \bar{\mathbf{T}} \bar{\mathbf{V}} \right)^*$, $\mathbf{C}_{k,l}^\text{W} = \sum\nolimits_{k^\prime \in \mathcal{K}_U} \text{diag} \left( \bar{\mathbf{Y}}_l \bar{\mathbf{T}} \mathbf{v}_{k^\prime} \right)^H \bar{\mathbf{X}}_l^H \bar{\boldsymbol{\Psi}}_k \bar{\mathbf{X}}_l \text{diag} \left( \bar{\mathbf{Y}}_l \bar{\mathbf{T}} \mathbf{v}_{k^\prime} \right)$, and the Hadamard product $\odot$.
Here, we have defined $\bar{\mathbf{V}} \bar{\mathbf{V}}^H = \sum\nolimits_{k\in\mathcal{K}_U} \mathbf{v}_k \mathbf{v}_k^H$, $\bar{\mathbf{X}}_l = \text{blkdiag} \left( \{ \mathbf{X}_{i,l} \}_{i\in\mathcal{K}_A} \right)$, and $\bar{\mathbf{Y}}_l = \text{blkdiag} \left( \{ \mathbf{Y}_{i,l} \}_{i\in\mathcal{K}_A} \right)$
with the matrices $\mathbf{X}_{i,l}$ and $\mathbf{Y}_{i,l}$ given by
\begin{subequations} 
\begin{align}
    &\!\!\!\!\mathbf{X}_{i,l} \triangleq 
    \begin{cases}
         \mathbf{\Phi}_{i,L} \mathbf{W}_{i,L} \mathbf{\Phi}_{i,L-1} \cdots \mathbf{\Phi}_{i,l+1} \mathbf{W}_{i,l+1}, & \!\!\!\text{if} \,\,l \neq L, \\
        \mathbf{I}_M, & \!\!\!\text{if} \,\,l=L.
    \end{cases}  
    \nonumber \\
    &\!\!\!\!\mathbf{Y}_{i,l} \triangleq 
    \begin{cases}
        \mathbf{W}_{i,l} \mathbf{\Phi}_{i,l-1} \cdots \mathbf{\Phi}_{i,2} \mathbf{W}_{i,2} \mathbf{\Phi}_{i,1}, & \!\!\!\text{if} \,\,l \neq 1, \\
        \mathbf{I}_M, & \!\!\!\text{if} \,\,l=1,
    \end{cases}  
    \nonumber
\end{align}
\end{subequations}

To handle the non-convex unit-modulus constraint (\ref{eq:problem-wave-unit-modulus-low-CQ}), we introduce an auxiliary variable $\boldsymbol{\phi}_l = [\boldsymbol{\phi}_{1,l}^H \, \cdots \, \boldsymbol{\phi}_{K_A,l}^H]^H \in \mathbb{C}^{MK_A \times 1}$, where $\boldsymbol{\phi}_{i,l} \in \mathbb{C}^{M\times 1}$ is used to encourage $\boldsymbol{\varphi}_{i,l}$ to satisfy the modulus constraint.
With this auxiliary variable, problem (\ref{eq:problem-wave-low}) can be equivalently reformulated as
\begin{subequations} \label{eq:problem-wave-low-penalty}
\begin{align}
    \!\!\!\!\underset{\boldsymbol{\varphi}_{l}, \boldsymbol{\phi}_l} {\mathrm{min.}}\,\,\, & \boldsymbol{\varphi}_{l}^H \mathbf{Q}_l^\text{W} \boldsymbol{\varphi}_{l} + 2\text{Re}\Big\{ \!\left(\mathbf{b}_l^\text{W}\right)^H \boldsymbol{\varphi}_{l} \Big\} + \xi\| \boldsymbol{\varphi}_{l} - \boldsymbol{\phi}_l \|^2 \, \label{eq:problem-wave-objective-low-penalty} \\
 \mathrm{s.t. }\quad & |\boldsymbol{\phi}_{i,l}(m)| = 1,\, \forall (i, m) \in \mathcal{K}_A \times \mathcal{M}, \label{eq:problem-wave-unit-modulus-low-penalty} 
\end{align}
\end{subequations}
where $\xi > 0$ is a penalty parameter controlling how strongly $\boldsymbol{\varphi}_{i,l}$ is driven to satisfy the unit-modulus constraint.

Since jointly optimizing $\boldsymbol{\varphi}_{i,l}$ and $\boldsymbol{\phi}_l$ remains difficult, we adopt an AO approach.
When $\boldsymbol{\phi}_l$ is kept fixed, problem (\ref{eq:problem-wave-low-penalty}) reduces to a convex quadratic problem whose optimal solution is given in closed form as
\begin{align}
    \boldsymbol{\varphi}_l = - ( \tilde{\mathbf{Q}}_l^\text{W} )^{-1} \tilde{\mathbf{b}}_l^\text{W}, \label{eq:wave-closed-form}
\end{align} 
with $\tilde{\mathbf{Q}}_l^\text{W} = \mathbf{Q}_l^\text{W} + \xi \mathbf{I}_{MK_A}$ and $\tilde{\mathbf{b}}_l^\text{W} = \mathbf{b}_l^\text{W} - \xi \boldsymbol{\phi}_l$.

Conversely, when $\boldsymbol{\varphi}_l$ is fixed, problem (\ref{eq:problem-wave-low-penalty}) simplifies to
    \begin{align}
    \!\!\!\!\underset{\boldsymbol{\phi}_l} {\mathrm{min.}}\,\,\, & \| \boldsymbol{\varphi}_{l} - \boldsymbol{\phi}_l \|^2 \label{eq:problem-phi-objective} \,\,\,\,
    \mathrm{s.t. } \,\, \mathrm{(\ref{eq:problem-wave-unit-modulus-low-penalty})},
\end{align}
which admits the closed-form solution $\boldsymbol{\phi}_l = \exp (j \angle \boldsymbol{\varphi}_{l} )$.

\subsection{Algorithm Description and Complexity Discussion} \label{sub:algorithm-complexity}

\begin{algorithm}
\caption{Proposed efficient AO algorithm}\label{algorithm}
\begin{algorithmic}[1]
\State \textbf{initialize:} 
\State Set $\mathbf{v}$ so that the constraint (\ref{eq:power-constraint}) is satisfied, and initialize the phase variables $\boldsymbol{\theta}$ within $[ 0, 2\pi)$ and the outer iteration count $n^{\text{out}}\leftarrow 1$.
    \Repeat
        \Repeat
            \State Update $\{\boldsymbol{\tau},\boldsymbol{\omega}\}$ with (\ref{eq:opt-auxiliary-tau-omega}).
            \State \textbf{for} $i \in \mathcal{K}_A$ \textbf{do}
                \State \quad\, Update $\mathbf{v}_{A,i}$ with (\ref{eq:lagrange-digital-opt}).
            \State \textbf{end}
        \Until {Converged or $n^{\text{digital}} \geq n_{\max}^{\text{digital}}$ (Otherwise, set 
        $n^{\text{digital}} \leftarrow n^{\text{digital}}+1$)}
        
        \State \textbf{for} $l \in \mathcal{L}$ \textbf{do}
            \State \quad\,\, Set $\xi \leftarrow \xi_0$.
            \State \quad\,\, Update $\{\boldsymbol{\tau},\boldsymbol{\omega}\}$ with (\ref{eq:opt-auxiliary-tau-omega}).
                \State \quad\,\, \textbf{repeat}
                    \State \qquad\quad Update $\boldsymbol{\phi}_l = \exp (j \angle \boldsymbol{\varphi}_{l} )$.
                    \State \qquad\quad Update $\boldsymbol{\varphi}_l$ with (\ref{eq:wave-closed-form}). \\
                    \qquad Update $\xi \leftarrow \varrho\xi$ $(\varrho>1)$.
                \State \quad\,\, \textbf{Until} {Converged or $n^{\text{wave}} \geq n_{\max}^{\text{wave}}$} (Otherwise, set $n^{\text{wave}} \leftarrow n^{\text{wave}}+1$)
        \State \textbf{end}
    \Until {Converged or $n^{\text{out}} \geq n_{\max}^{\text{out}}$ (Otherwise, set $n^{\text{out}} \leftarrow n^{\text{out}}+1$)}
\end{algorithmic}
\end{algorithm}

The proposed AO algorithm for the joint optimization of the digital beamforming $\mathbf{v}$ and the wave-domain beamforming $\boldsymbol{\theta}$ is summarized in Algorithm 1. 
Specifically, the digital and wave-domain beamforming variables are alternately optimized in Steps 4--9 and 10--18, respectively.

The overall computational complexity $C$ of Algorithm 1 can be expressed as $C = I_{\text{out}} ( C_{\text{digital}} + C_{\text{wave}} )$, where
$C_{\text{digital}}$ and $C_{\text{wave}}$ denote the complexities associated with the digital and wave-domain beamforming updates, respectively, and $I_{\text{out}}$ represents the number of outer iterations required for convergence.
The complexity $C_{\text{digital}}$ of the digital beamforming update is proportional to the per-inner-iteration complexity of Steps 5--8, which scales as $\mathcal{O}(K_AK_UN ( K_AK_U + K_U^2N^2 + M^2 ))$. This complexity is significantly lower than the worst-case complexity $\mathcal{O}(K_A^4K_U^4N^4)$ incurred when CVX is used to solve (\ref{eq:problem-convexified-digital}) while keeping $\{\boldsymbol{\tau}, \boldsymbol{\omega}\}$ fixed \cite[p. 4]{BTal:LN19}.
The complexity $C_{\text{wave}}$ of the wave-domain beamforming update linearly increases with the number of SIM layers $L$.
For each layer, the dominant computational cost arises from updating the phase-shift variables $\boldsymbol{\varphi}_l$ in Step 15, whose complexity scales as $\mathcal{O}(K_AM^2 (K_A^2K_UN + K_U^2M + K_A^2M))$. 

\section{Numerical Results} \label{sec:numerical}


We consider a hexagonal coverage area with a radius of 100 m \cite[Fig. 1]{Park:CISS14}, where randomly distributed UEs are served by SIM-equipped APs located at equi-spaced boundary points and employing sectorized antennas directed toward the center.
Each AP is equipped with $N=2$ RF chains.
Unless stated otherwise, each SIM layer consists of $M=16$ meta-atoms arranged in a $4\times 4$ uniform planar array.
The pathloss is modeled as $\beta_{k, i} = \beta_0 ( d_{k,i}^{\text{geo}}/d_0 )^{-3}$, where $d_{k,i}^{\text{geo}}$ is the distance between UE $k$ and AP $i$.
The reference distance and pathloss are set to $d_0 = 1$ m and $\beta_0 = (\lambda/4\pi d_0)^2$, respectively. 
The carrier frequency is 28 GHz, and the noise power is $\sigma_z^2=-104$ dBm.
The thickness of each SIM is $T_{\text{SIM}} = 5\lambda$, and the inter-layer spacing is $d_{\text{Layer}} = T_{\text{SIM}}/L$.
The area of each meta-atom is given by $S = (\lambda / 2)^2$.
We evaluate the unweighted sum-rate performance throughout the simulations.
The covariance matrix $\boldsymbol{\Psi}_{k,i}$ of the channel error vector $\mathbf{e}_{k,i}$ is modeled as $\boldsymbol{\Psi}_{k,i} = \rho\beta_{k,i}\mathbf{R}_{k,i}$, where the estimation error coefficient $\rho \in [0,1]$ quantifies the normalized level of CSI uncertainty. 


We compare the proposed algorithm with the following AO-based benchmark schemes: 1) \textit{CVX-GA}: The optimization of $\mathbf{v}$ in (\ref{eq:problem-convexified-digital}), while keeping $\{\boldsymbol{\tau}, \boldsymbol{\omega}\}$ fixed, is performed using CVX;
2) \textit{Wave-only}: Digital beamforming is restricted to power control, i.e., $\mathbf{v}_{k,i}\in\mathbb{R}^{N \times 1}_+$; 3) \textit{MRT (frac.)}: Digital beamforming follows maximum ratio transmission (MRT) with fractional power allocation \cite{Hu:TVT25}, i.e., $\mathbf{v}_{k,i} = \sqrt{P_{\text{tx}}\beta_{k,i}/({N\sum_{k\in\mathcal{K}_U}\beta_{k,i}})} \tilde{\mathbf{h}}_{k,i}/ \|\tilde{\mathbf{h}}_{k,i}\|$; 4) \textit{MRT (eq.)}: MRT beamforming with equal power allocation; and 5) \textit{Rand. wave}: The phase shifts $\boldsymbol{\theta}$ are randomly fixed, and only the digital beamforming variables are optimized via Steps 4--9 of Algorithm 1.
Except for the random wave scheme, the wave-domain beamforming $\boldsymbol{\theta}$ is optimized using a gradient ascent (GA) method for given digital beamforming variables.

\begin{figure}
\centering\includegraphics[width=0.75\linewidth]{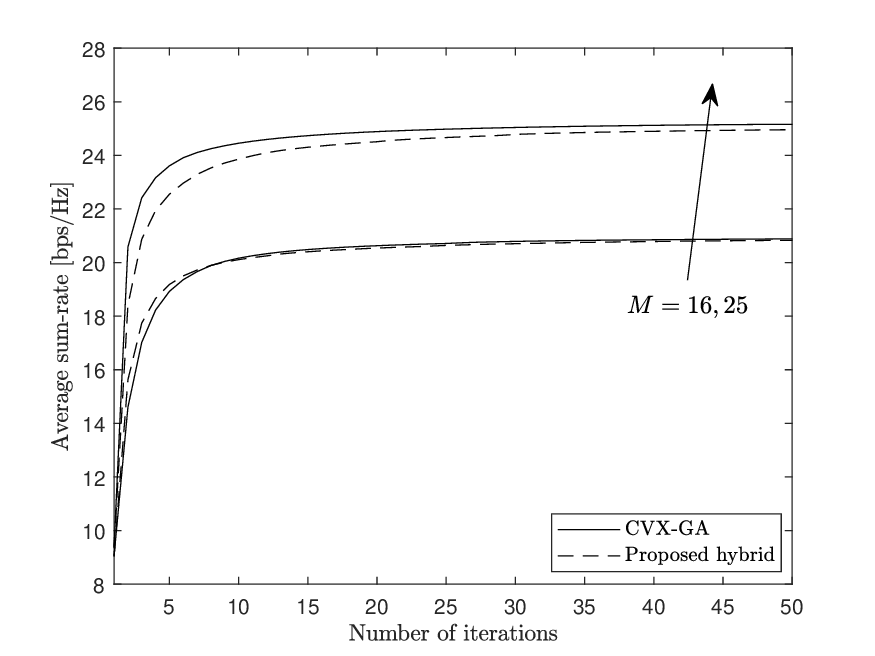}
\vspace{-2mm}
\caption{\small Average sum-rate versus the number of iterations ($K_A=3$, $K_U=6$,  $P_\text{tx}=30$ dBm, $L=4$, $M\in\{16,25\}$ and $\rho=0.1$).} \label{fig:graph-vs-iterations}
\vspace{-2mm}
\end{figure}

Fig. \ref{fig:graph-vs-iterations} plots the average sum-rate versus the number of iterations for $K_A=3$, $K_U=6$,  $P_\text{tx}=30$ dBm, $L=4$, $M\in\{16,25\}$ and $\rho=0.1$. The figure shows that both the proposed and CVX-GA schemes achieve monotonically increasing sum-rates and converge within a few iterations across all simulated configurations, reaching nearly identical performance levels.

\begin{figure}
\centering\includegraphics[width=0.75\linewidth]{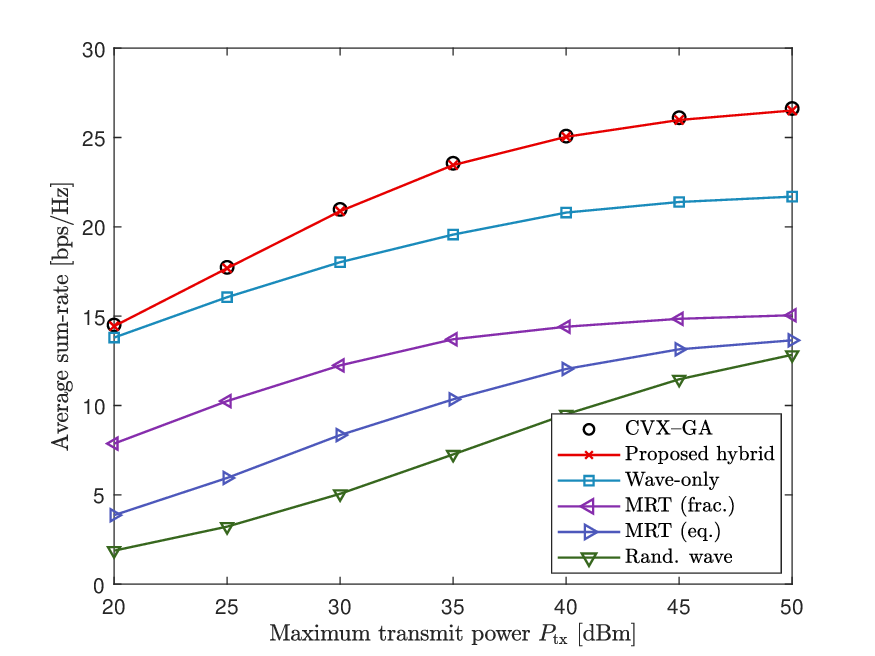}
\vspace{-2mm}
\caption{\small Average sum-rate versus the maximum transmit power $P_\text{tx}$ ($K_A=3$, $K_U=6$, $L=4$, $M=16$ and $\rho=0.1$).} \label{fig:graph-vs-P}
\vspace{-2mm}
\end{figure}

Fig. \ref{fig:graph-vs-P} illustrates the average sum-rate as a function of the maximum transmit power $P_\text{tx}$ for $K_A=3$, $K_U=6$, $L=4$, $M=16$ and $\rho=0.1$.
The proposed scheme closely approaches the performance of the CVX-GA scheme across all simulated scenarios while requiring significantly lower computational complexity as will be demonstrated later in this section.
Moreover, it substantially outperforms the wave-only, MRT, and random-wave schemes, highlighting the benefit of jointly optimizing digital and wave-domain beamforming.
The performance gains increase with $P_{\text{tx}}$, since the impact of residual interference becomes more pronounced at higher $P_{\text{tx}}$.

\begin{figure}
\centering\includegraphics[width=0.75\linewidth]{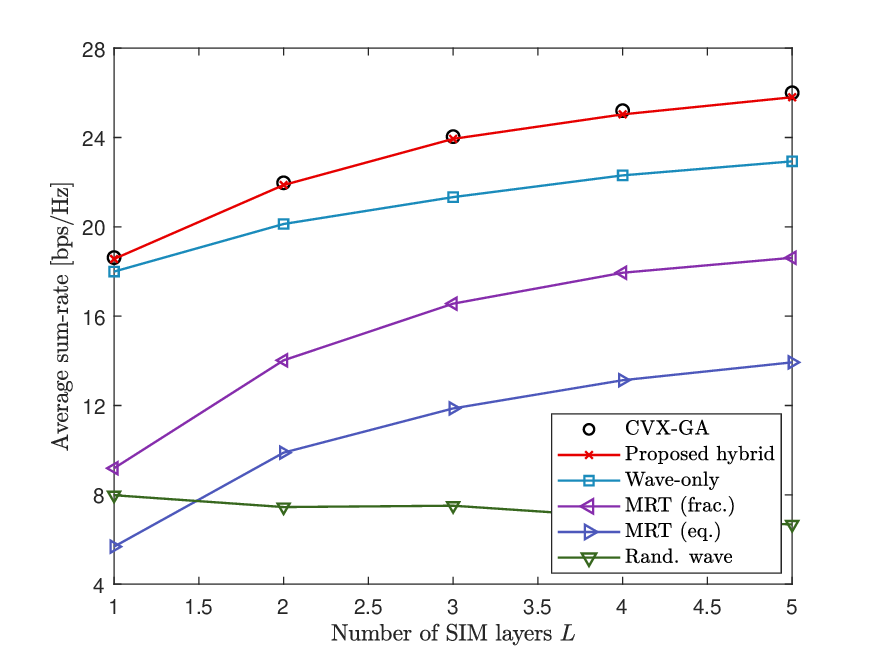}
\vspace{-2mm}
\caption{\small Average sum-rate versus the number of metasurface layers $L$ ($K_A=3$, $K_U=6$, $P_\text{tx}=30$ dBm, $L=2$, $M=25$ and $\rho=0.1$).} \label{fig:graph-vs-layer}
\vspace{-2mm}
\end{figure}

Fig. \ref{fig:graph-vs-layer} presents the average sum-rate versus the number of metasurface layers $L$ for $K_A=3$, $K_U=6$, $P_\text{tx}=30$ dBm, $L=2$, $M=25$ and $\rho=0.1$.
Similar to Fig. \ref{fig:graph-vs-P}, the proposed scheme achieves nearly identical performance to the CVX-GA scheme while maintaining significantly lower computational complexity.
As the number of SIM layers increases, most schemes achieve improved performance due to the higher degrees of control provided by the SIM phase shifts.
In contrast, the random-wave scheme does not benefit from this increase since the phase shifts are randomly fixed.

\begin{figure}
\centering\includegraphics[width=0.75\linewidth]{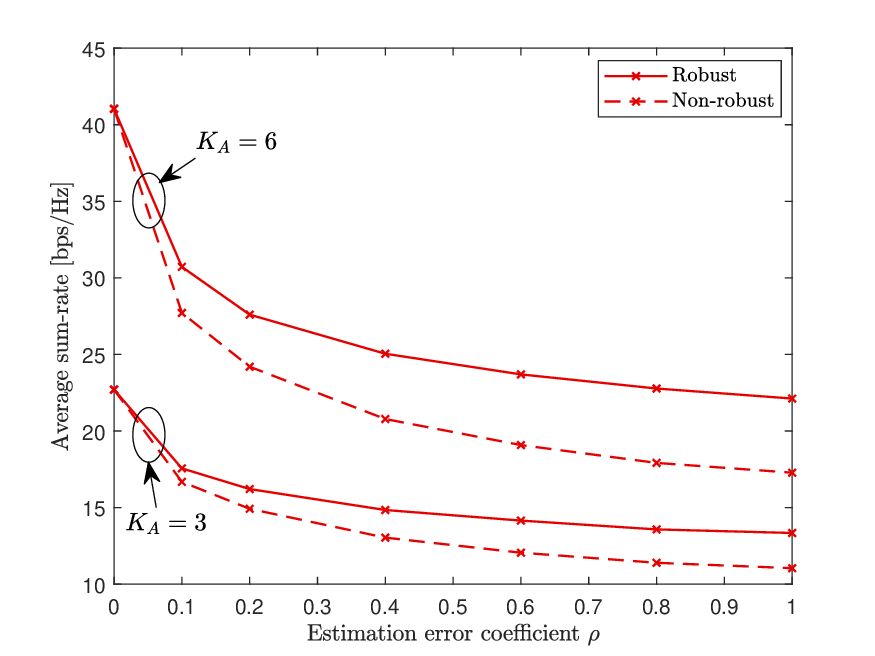}
\vspace{-2mm}
\caption{\small Average sum-rate versus the channel estimation error coefficient $\rho$ ($K_A\in\{3,6\}$, $K_U=6$, $P_\text{tx}=30$ dBm, $L=2$ and $M=16$).} \label{fig:graph-vs-rho}
\vspace{-2mm}
\end{figure}

In Fig. \ref{fig:graph-vs-rho}, we plot the average sum-rate versus the channel estimation error coefficient $\rho$ for $K_A\in\{3,6\}$, $K_U=6$,  $P_\text{tx}=30$ dBm, $L=2$ and $M=16$. The non-robust scheme optimizes the beamforming variables assuming that the estimated channels do not contain any errors.
As expected, the performance gap between the proposed robust scheme and the non-robust benchmark widens as $\rho$ increases due to the growing impact of CSI uncertainty. Furthermore, the robustness of the proposed scheme becomes more pronounced as the number of APs $K_A$ increases, since larger distributed arrays are more sensitive to CSI mismatches.

\begin{figure}
\centering\includegraphics[width=0.75\linewidth]{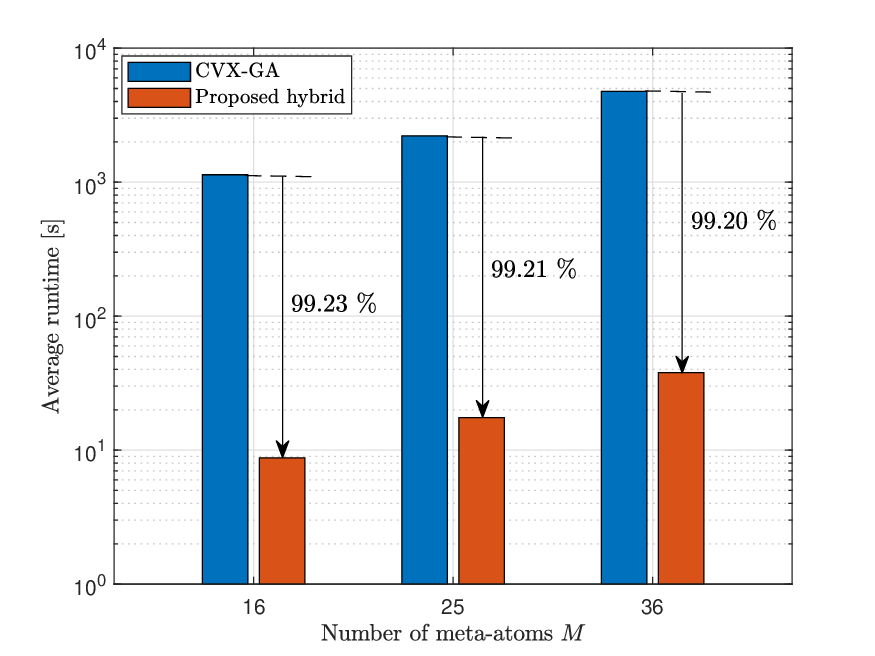}
\vspace{-2mm}
\caption{\small Average runtime versus the number of meta-atoms $M$ ($K_A=3$, $K_U=6$, $P_\text{tx}=30$ dBm, $L=4$ and $\rho=0.1$).} \label{fig:graph-vs-runtime}
\vspace{-2mm}
\end{figure}

Fig. \ref{fig:graph-vs-runtime} compares the average runtime of the proposed hybrid scheme and the CVX-GA scheme for $K_A=3$, $K_U=6$,  $P_\text{tx}=30$ dBm, $L=4$, and $\rho=0.1$. While we have observed through Figs. \ref{fig:graph-vs-iterations}--\ref{fig:graph-vs-layer} that both schemes achieve nearly identical sum-rate performance, the proposed scheme reduces the runtime by more than 99\%, demonstrating its practical advantage for large-scale deployments.

\section{Conclusion} \label{sec:conclusion}
In this paper, we have developed an efficient optimization algorithm for hybrid digital and wave-domain beamforming in SIM-assisted CF-mMIMO systems.
To address the prohibitive computational complexity of conventional AO frameworks, caused by the reliance on general-purpose solvers and per-element GA, the proposed AO algorithm updates both digital and wave-domain beamforming variables in closed form.
Numerical results have demonstrated that the proposed scheme achieves more than 99\% complexity reduction while attaining nearly identical sum-rate performance compared with the conventional AO approach.
As future work, the proposed framework can be extended to more practical scenarios accounting for finite-capacity fronthaul links, discrete SIM phase shifts.

\end{document}